\documentclass[prb,twocolumn,showpacs]{revtex4}
\usepackage{graphicx,amsmath,epsf}

\begin{document}
\title{Semiclassical theory of a quantum pump}

\author{Saar Rahav}
\affiliation{Laboratory of Atomic and Solid State Physics, Cornell University, Ithaca 14853, USA.}
\author{Piet W. Brouwer}
\affiliation{Laboratory of Atomic and Solid State Physics, Cornell University, Ithaca 14853, USA.}

\date{\today}

\begin{abstract}
In a quantum charge pump, the periodic variation of two parameters that
affect the phase of the electronic wavefunction causes the flow
of a direct current. The operating mechanism of a quantum pump is
based on quantum interference, the phases of interfering amplitudes
being modulated by the external parameters. In a ballistic quantum
dot, there is a minimum time before which quantum interference can not
occur: the Ehrenfest time $\tau_{\rm E}$. Here we calculate the current pumped
through a ballistic quantum dot if $\tau_{\rm E}$ is comparable to
the mean dwell time $\tau_{\rm D}$.
Remarkably, we find that the pumped current has a
component that is not suppressed if $\tau_{\rm E} \gg \tau_{\rm
  D}$.
\pacs{73.23.-b,05.45.Mt,73.50.Pz}
\end{abstract}
\maketitle

\section{Introduction}

A suitably chosen
periodic perturbation of an electronic device may result in a direct
current, even when the system is not biased. In the adiabatic limit,
when the applied perturbation is slow in
comparison to the escape rate to external contacts, the electronic state
is the same after each period while a finite charge is transferred
through the device during each cycle. Such devices are called
charge pumps.
Charge pumps have been proposed as current sources and
minimal-noise current
standards,\cite{kn:pothier1992,kn:martinis1994,kn:avron2001} as well as
diagnostic tools to
monitor how mesoscopic devices respond to changing external
parameters.\cite{kn:thouless1983,kn:falko1989,kn:bykov1989,kn:niu1990,kn:spivak1995,kn:switkes1999,kn:ebbecke2004}

In this article, we consider a so-called `quantum charge
pump'. In a quantum
pump, the external perturbation affects the phases of the electron
wavefunctions only, not the classical dynamics of the
electrons.\cite{kn:switkes1999,kn:thouless1983,kn:falko1989,kn:spivak1995,kn:brouwer1998,kn:altshuler1999,kn:zhou1999}
Such a scenario can be
realized in a ballistic quantum dot, where small variations
of shape-defining gate voltages or of an applied magnetic field are
known to significantly affect the electronic phase while leaving
classical trajectories unaffected. Such a device, with two
shape-distorting gate voltages to drive the current, was built by Switkes
{\em et al.}\cite{kn:switkes1999}
Although the pumped current in the original experiment of Ref.\
\onlinecite{kn:switkes1999} was obscured by rectification
effects,\cite{kn:brouwer2001} `quantum pumping'
was demonstrated in a later experiment, be
it not in the adiabatic regime.\cite{kn:dicarlo2003}

Microscopically, the current in a quantum pump is the result of
quantum interference: two classical trajectories initially a
microscopic distance apart (a Fermi wavelength) split and rejoin,
with a phase difference that depends on the applied perturbation.
Indeed, if the mean dwell time $\tau_{\rm D}$ in the device exceeds
the dephasing time $\tau_{\phi}$, the pumped current is
suppressed.\cite{kn:cremers2002,kn:moskalets2001} Aleiner and Larkin
have pointed out that, in a ballistic quantum dot (as in any quantum
system with a well-defined classical counterpart for which the
dynamics is chaotic), there is another time that governs quantum
interference:\cite{kn:aleiner1996,kn:aleiner1997} the Ehrenfest time
$\tau_{\rm E}$. The Ehrenfest time is the time during which two
classical trajectories initially a Fermi wavelength apart are
separated to a macroscopic distance (the width of the contacts or
the size of the quantum dot) by the chaotic classical
dynamics.\cite{kn:zaslavsky1981,kn:larkin1968} The Ehrenfest time is
given by
\begin{equation}
  \tau_{\rm E} = \frac{1}{\lambda} \ln N + \mbox{const},
  \label{eq:tauE}
\end{equation}
where $\lambda$ is the Lyapunov exponent of the classical dynamics in
the quantum dot and $N$ denotes the total number of
channels in the contacts. The constant in Eq.\
(\ref{eq:tauE}) is independent of $N$.
While the dephasing time governs the suppression of interference for
long dwell times, the Ehrenfest time is the minimum time
needed for the appearance of interference effects.

Being an interference effect, the
current pumped through a ballistic chaotic quantum dot is a random
function of any parameter that affects quantum phases, such as
the shape of the dot or the applied magnetic field (taken at
a reference point during the cycle). Therefore,
one typically considers averages and fluctuations of the current,
taken with respect to an ensemble of dots with slightly varying shape
or with respect to a range of magnetic field strengths. Since the
ensemble-averaged current $\langle I \rangle = 0$, the magnitude of the
pumped current is measured through the second moment $\langle I^2
\rangle$. With a few
exceptions,\cite{kn:zhou1999,kn:martinez-mares2004}
random matrix theory has been the preferred
framework for a statistical theory of quantum-dot based charge
pumps.\cite{kn:brouwer1998,kn:shutenko2000,kn:vavilov2001a,kn:polianski2002,kn:polianski2003,kn:vavilov2005}
Random matrix
theory describes the regime $\tau_{\rm E} \ll \tau_{\rm D}$, in which
quantum effects occur on a time scale much shorter than the mean dwell
time $\tau_{\rm D}$ in the quantum dot.
In this article, we are interested in a quantum pump in
the regime in which
$\tau_{\rm E}$ and $\tau_{\rm D}$ are comparable.  Since, typically,
$\tau_{\rm D} \lambda \gg 1$, this requires that we consider the
classical limit $N \gg 1$.

A remarkably wide
range of $\tau_{\rm E}$-dependences has been reported in the
literature for various quantum interference effects in ballistic
quantum dots. Weak localization, the quantum interference correction
to the
ensemble-averaged conductance in the presence of time-reversal
symmetry, is suppressed $\propto \exp(-\tau_{\rm
  E}/\tau_{\rm D})$ if $\tau_{\rm E}$ is
large.\cite{kn:aleiner1996,kn:adagideli2003,kn:rahav2005,kn:jacquod2006}
On the other hand, conductance fluctuations have no $\tau_{\rm E}$
dependence,\cite{kn:tworzydlo2004,kn:jacquod2004,kn:brouwer2006}
whereas the quantum correction to the spectral form factor of a closed
quantum dot with broken time-reversal symmetry exists for finite
$\tau_{\rm E}$ only,\cite{kn:tian2004b,kn:brouwer2006b} being absent in the random
matrix limit $\tau_{\rm E} \to 0$.\cite{kn:berry1985} The main result
of the present article is that current
pumped in a quantum pump has yet another $\tau_{\rm E}$-dependence,
\begin{equation}
  \langle I^2 \rangle  = \frac{1}{2}
  (1 + e^{-2
  \tau_{\rm E}/\tau_{\rm D}}) \langle I^2 \rangle_{\rm RMT},
  \label{eq:Ifinal}
\end{equation}
where $\langle I^2 \rangle_{\rm RMT}$ is the prediction of
random matrix theory.

We calculate the Ehrenfest-time dependence of the pumped current
using a semiclassical theory in which the pumped current is expressed
as a sum over classical trajectories in the quantum dot. Our theory
goes beyond
a previous semiclassical theory of quantum pumps by Martinez-Mares
{\em et al.},\cite{kn:martinez-mares2004}
who did not consider the role of the Ehrenfest time.
Whereas Ref.\ \onlinecite{kn:martinez-mares2004} calculated the pumped
current in the `diagonal approximation', in which only pairs of
identical classical trajectories are considered, we also include the
leading
off-diagonal terms in the sum over classical trajectories. In this
respect, our theory builds
on previous work by Richter and Sieber\cite{kn:richter2002} and
Heusler {\em et al.},\cite{kn:heusler2006} who developed a systematic
way to include off-diagonal terms in trajectory sums of the weak
localization correction to the dot's conductance
and other quantum
interference corrections to transport. Unlike the diagonal
approximation, which is known to violate current conservation, the
present version of the semiclassical theory is fully current
conserving.


In Sec.\ \ref{pumps} we express the pumped current in terms of a sum
over classical trajectories. This section closely follows Ref.\
\onlinecite{kn:martinez-mares2004}. The sum over classical trajectories,
which gives the final result (\ref{eq:Ifinal}), is then performed in
Sec.\ \ref{semic}. We compare the theoretical predictions to
numerical simulations in Sec.\ \ref{numerics} and conclude in Sec.\
\ref{disc}.

\section{Semiclassical theory of a quantum pump}
\label{pumps}

\begin{figure}[t]
\epsfxsize=0.9\hsize
\hspace{0.01\hsize}
\epsffile{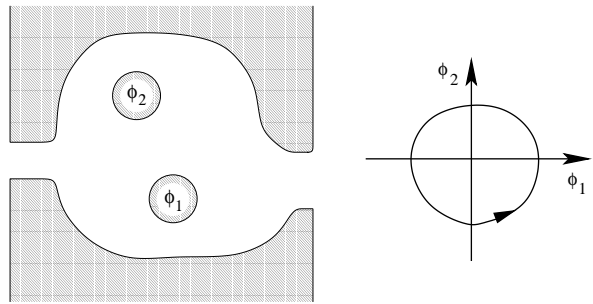}
\caption{\label{dot1} Left: Schematic drawing of the quantum pump under
  consideration. The quantum pump consists of a ballistic quantum dot
  with two holes, through which a time-dependent magnetic flux $\phi$
  is applied. The classical dynamics in the dot is assumed to be chaotic. Right: pumping cycle in flux-space.}
\end{figure}

The geometry under consideration is shown in Fig.\ \ref{dot1}. For
definiteness, we consider a ballistic chaotic quantum dot coupled to
two electron reservoirs via ballistic point contacts. The quantum dot
has two holes. Slow periodic variations of the magnetic flux through
each of the holes change the phases of wavefunctions, not the
classical trajectories. Both holes are `macroscopic': their size is
large in comparison to the Fermi wavelength $\lambda_F$
and their boundary is smooth.

We assume that the variation of the fluxes is adiabatic, with period
$2 \pi/\omega$
much larger than either the dwell time or the Ehrenfest time. In the
adiabatic regime, the time-averaged current $I_L$ through the left
contact can be expressed in terms of the $N \times N$
scattering matrix $S$ of the quantum dot,\cite{kn:brouwer1998}
\begin{eqnarray}
  I_L & =& \frac{e \omega}{2 \pi} \int_{{\cal A}} d \phi_1 d \phi_2
  \Pi^{L} (\phi_1, \phi_2), \nonumber \\
  \Pi^{L} (\phi_1, \phi_2) & = & \frac{1}{\pi} \mbox{Im}\,
  \sum_{\nu=1}^{N}
  \sum_{\mu=1}^{N_1}
  \frac{\partial S^{*}_{\mu \nu}}{\partial
  \phi_1} \frac{\partial S_{\mu \nu}}{\partial \phi_2}.
  \label{eq:Ipump}
\end{eqnarray}
Here $\phi_j$ is the flux through hole $j$, $j=1,2$, measured in units
of the flux quantum $hc/e$, and the integral in Eq.\ (\ref{eq:Ipump})
is taken over the area enclosed in $(\phi_1,\phi_2)$-space in one
pumping cycle, see Fig.\ \ref{dot1}.
The number of channels in the left and
right contacts are $N_1$ and $N_2$, respectively, and $N = N_1 +
N_2$. In Eq.\ (\ref{eq:Ipump}) we assumed zero temperature. The
current through the right-contact reads
\begin{eqnarray}
  I_R & =& \frac{e \omega}{2 \pi} \int_{{\cal A}} d \phi_1 d \phi_2
  \Pi^{R} (\phi_1, \phi_2), \nonumber \\
  \Pi^{R} (\phi_1, \phi_2) & = & \frac{1}{\pi} \mbox{Im}\, \sum_{\nu=1}^{N}
  \sum_{\mu=N_1+1}^{N}
  \frac{\partial S^{*}_{\mu \nu}}{\partial
  \phi_1} \frac{\partial S_{\mu \nu}}{\partial \phi_2}.
  \label{eq:IpumpR}
\end{eqnarray}
Current conservation implies
\begin{equation}
  I = I_L = - I_R,
\end{equation}
so that
\begin{equation}
  \Pi^L = - \Pi^R.
  \label{eq:unitary}
\end{equation}

The key point of the semiclassical approach is to express the
scattering matrix $S$ as a sum over classical
trajectories $\gamma$,\cite{kn:jalabert1990,kn:baranger1993,kn:baranger1993b}
\begin{equation}
  S_{\mu\nu} = \frac{1}{\sqrt{N\tau_{\rm D}}} \sum_{\gamma}
  A_\gamma e^{i \frac{{\cal S}_\gamma}{\hbar}},
  \label{eq:Ssemi}
\end{equation}
where $\gamma$ starts at a contact with transverse momentum compatible
with lead mode $\nu$ and ends with transverse momentum compatible
with mode $\mu$. (The modes $\mu=1,\ldots,N_1$ are in the
left lead and the modes $\mu=N_1+1,\ldots,N$ are in the right lead.)
Further ${\cal S}_{\gamma}$ is the classical
action (which includes Maslov phases) and
$A_{\gamma}$ is the stability amplitude. The classical action
is modified by the presence of the fluxes $\phi_1$ and $\phi_2$ as
\begin{equation}
  {\cal S}(\gamma) = {\cal S}_0(\gamma) + h [\phi_1 W_1(\gamma) +
  \phi_2 W_2(\gamma)],
  \label{eq:action}
\end{equation}
where $W_j(\gamma)$ is the winding number of the trajectory $\gamma$
around the hole $j$, $j=1,2$. We'll be interested in the regime $\phi_j
\ll 1$, and $W_j \gg 1$, for which the discreteness of the winding
numbers does not play a role. Substituting Eqs.\ (\ref{eq:Ssemi}) and
(\ref{eq:action}) into Eqs.\ (\ref{eq:Ipump}) and (\ref{eq:IpumpR}),
we arrive at\cite{kn:martinez-mares2004}
\begin{eqnarray}
  \Pi^L &=& \frac{4 \pi}{N \tau_{\rm D}}
  \sum_{\nu=1}^{N} \sum_{\mu=1}^{N_1} \sum_{\gamma_1,\gamma_2}
  A_{\gamma_1} A_{\gamma_2} W_1(\gamma_1) W_2(\gamma_2)
  \nonumber \\ && \mbox{} \times
  \sin[({\cal S}_{\gamma_2} - {\cal S}_{\gamma_1})/\hbar],
  \nonumber \\
  \Pi^R &=& \frac{4 \pi}{N \tau_{\rm D}}
  \sum_{\nu=1}^{N} \sum_{\mu=N_1+1}^{N} \sum_{\gamma_1,\gamma_2}
  A_{\gamma_1} A_{\gamma_2} W_1(\gamma_1) W_2(\gamma_2)
  \nonumber \\ && \mbox{} \times
  \sin[({\cal S}_{\gamma_2} - {\cal S}_{\gamma_1})/\hbar],
  \label{eq:Isemi}
\end{eqnarray}
where both $\gamma_1$ and $\gamma_2$ are compatible with modes
$\nu$ and $\mu$ upon entrance and exit, respectively.

In the remainder of this article we limit ourselves to
bilinear response, $I \propto {\cal A}$. For bilinear response,
the fluctuations of the kernels $\Pi^L$
and $\Pi^R$ can be neglected when performing the integral over the
enclosed area ${\cal A}$ in $(\phi_1,\phi_2)$-space, so that
it is sufficient to calculate the average and variance of
the kernels $\Pi^L$ and $\Pi^R$ in order to find the average and
variance of the pumped current $I$,
\begin{eqnarray}
  \langle I \rangle &=& \frac{e \omega {\cal A}}{2 \pi} \langle \Pi^L \rangle
  = - \frac{e \omega {\cal A}}{2 \pi} \langle \Pi^R \rangle \\
  \langle I^2 \rangle &=& \frac{e^2 \omega^2 {\cal A}^2}{4 \pi^2}
  \langle (\Pi^L)^2 \rangle = 
  - \frac{e^2 \omega^2 {\cal A}^2}{4 \pi^2}
  \langle \Pi^L \Pi^R \rangle.
  \label{eq:varI}
\end{eqnarray}
In the semiclassical framework, this means that we can neglect the
dependence of the classical actions ${\cal S}_{\gamma_j}$ on the
variations of the parameters $\phi_1$ and $\phi_2$ in the sines in
Eq.\ (\ref{eq:Isemi}).

\begin{figure}[t]
\epsfxsize=0.6\hsize
\hspace{0.01\hsize}
\epsffile{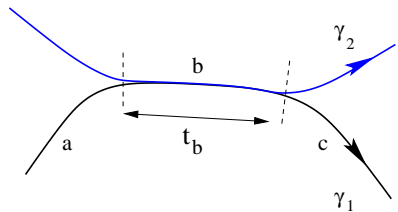}
\caption{(Color online)\label{corr} Schematic drawing of two partially correlated
  trajectories. The true trajectories are piecewise straight with
  specular reflection off the dot's boundaries.}
\end{figure}

Before we can perform the summation over classical trajectories necessary
to calculate $\langle I^2 \rangle$ we must specify the statistics
of the winding numbers $W_1$ and
$W_2$. For a quantum dot with chaotic classical dynamics, one may
assume that
the two winding numbers $W_1(\gamma)$ and $W_2(\gamma)$ are statistically
uncorrelated, and that the average winding number is zero,
\begin{equation}
  \langle W_j(\gamma) \rangle = 0,\ \ j=1,2.
  \label{eq:w1}
\end{equation}
For the
variance of the winding number we take\cite{kn:berry1986}
\begin{equation}
  \langle W_j(\gamma)^2 \rangle = C_j t,\ \ j=1,2,
  \label{eq:w}
\end{equation}
where the coefficients $C_j$ depend on the size and shape of the
quantum dot and on the location of the hole $j$, $j=1,2$,
but not on the Ehrenfest time and $t$ is the duration of the
trajectory $\gamma$.
For a chaotic quantum dot, one may also assume
that the winding numbers $W_j(\gamma_1)$ and $W_j(\gamma_2)$ of
different trajectories $\gamma_1$ and $\gamma_2$ are uncorrelated,
$j=1,2$,
except if there exist strong classical correlations between the two
trajectories. Strong classical correlations exist if the two
trajectories are within a phase space distance $c$, where $c$ is a
cut-off below which the chaotic classical dynamics in the quantum dot
can be linearized. If two trajectories are correlated for only a part
of their duration, as shown schematically in Fig.\ \ref{corr}, only
the duration $t_b$ of the correlated segment contributes to the
average of the product of the winding numbers,
\begin{equation}
  \langle W_j (\gamma_1) W_j(\gamma_2) \rangle = C_j t_b,\ \
  j=1,2.
\end{equation}

Although our theory is formulated for a quantum pump where the current
is driven by time-dependent magnetic fluxes, the mathematical
framework to describe other geometries, {\em e.g.} a pump in which
current is driven by shape changes, is
identical.\cite{kn:martinez-mares2004} For generic perturbations of a
chaotic quantum dot the parameters still enter the classical action
as a Gaussian random
process, and the actions of different trajectories will have
statistically uncorrelated dependences on the parameters. In that
sense, our theory applies to arbitrary perturbations, not only
time-dependent magnetic fluxes. (The only
exception is a pump in which a gate voltage is changed uniformly in the
quantum dot.
A similar
separation between `generic' perturbations and a uniform gate voltage
shift exists in the random matrix description of quantum
pumps.\cite{kn:brouwer1998,kn:vavilov2001a,kn:polianski2003})

\section{Summation over classical trajectories}
\label{semic}

The semiclassical framework and the statistics of the winding
numbers outlined in the previous section is the same as that used in
a previous semiclassical theory of quantum pumps by Martinez-Mares,
Lewenkopf, and Mucciolo.\cite{kn:martinez-mares2004} In Ref.\
\onlinecite{kn:martinez-mares2004}, the fourfold sum over classical
trajectories required to calculate the current variance is evaluated
in the `diagonal approximation', in which the four trajectories
contributing $\langle I^2 \rangle$ are taken pairwise equal.
Although Ref.\ \onlinecite{kn:martinez-mares2004} reports agreement
between semiclassics and random matrix theory, the use of the
diagonal approximation to calculate the current variance $\langle
I^2 \rangle$ is problematic because it violates the unitarity
relation (\ref{eq:unitary}): In the diagonal approximation, one has
$\langle \Pi^L \Pi^R \rangle = 0$, while $\langle (\Pi^L)^2 \rangle
\neq 0$. The diagonal approximation also fails to account for the
fact that the variance of the pumped current does not depend on the
presence or absence of time-reversal
symmetry.\cite{kn:brouwer1998,kn:shutenko2000}

Problems with the diagonal approximation are not limited to the
semiclassical theory of a quantum pump. In fact, the diagonal
approximation is well known to fail at providing a
current-conserving description of other quantum interference effects,
such as the weak localization correction to the conductance
and universal conductance fluctuations.\cite{kn:stone1995} For weak
localization, this problem was solved by Richter and Sieber, who were
able to include the relevant off-diagonal configurations of classical
trajectories into the trajectory sum.\cite{kn:richter2002} Below, we
use a generalization of Richter and Sieber's
technical
innovation\cite{kn:heusler2006,kn:braun2005,kn:brouwer2006,kn:brouwer2006c}
to perform the trajectory sums required for a
current-conserving semiclassical theory of a quantum pump.

\begin{figure}[t]
\epsfxsize=0.9\hsize
\hspace{0.01\hsize}
\epsffile{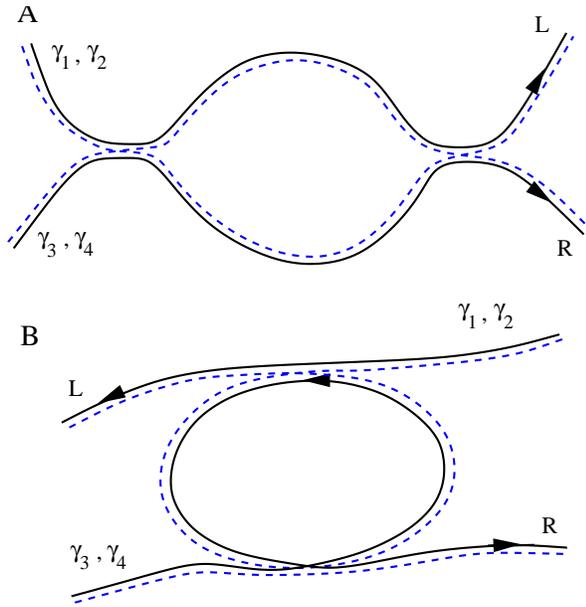}
%
\caption{(Color online)\label{typea}
Schematic drawing of trajectory configurations of types A (top) and B
(bottom) that generate the pumped current. Type A consists of all
quadruples of classical trajectories with two separated small-angle
encounters. Type B consist of all quadruples of classical trajectories
for which two (or more) of the trajectories revolve around a periodic
orbit. If the Ehrenfest time is much smaller than the mean dwell time,
the pumped current is carried by trajectories of type A. If $\tau_{\rm
E} \gg \tau_{\rm D}$, the current is carried by trajectories of type B.}
\end{figure}

Since the two winding numbers $W_1$ and $W_2$ are statistically
uncorrelated and have zero average, the trajectory
sum in Eq.\ (\ref{eq:Isemi}) immediately gives $\langle
\Pi^L \rangle = \langle \Pi^R \rangle = 0$, and hence
\begin{equation}
  \langle I \rangle = 0.
\end{equation}
In order to calculate the average of the squared current, it is
technically most convenient to use the second equality in Eq.\
(\ref{eq:varI}). Using the semiclassical expression (\ref{eq:Isemi}),
one has
\begin{eqnarray}
\label{fullsum}
  \left< \Pi^L \Pi^R \right> & = & \frac{16 \pi^2}{N^2 \tau^2_{\rm D}}
  \sum_{\mu,\rho=1}^{N} \sum_{\nu=1}^{N_1} \sum_{\sigma=N_1+1}^{N}
  \sum_{\gamma_1,\gamma_2,\gamma_3,\gamma_4}
  \nonumber \\ && \mbox{} \times
  \left\langle A_{\gamma_1} A_{\gamma_2} A_{\gamma_3} A_{\gamma_4}
  \right. \nonumber \\ && \left. \mbox{} \times
  W_1(\gamma_1) W_2 (\gamma_2) \sin[({{\cal S}_{\gamma_2} -{\cal S}_{\gamma_1}
  })/{\hbar}]
  \right. \nonumber \\ &&  \left. \mbox{} \times
  W_1(\gamma_3) W_2 (\gamma_4) \sin[({{\cal S}_{\gamma_4}
  -{\cal S}_{\gamma_3}})/{\hbar}]
  \right>. ~~~
\end{eqnarray}
The non-vanishing contributions to the current variance are from
quadruples of trajectories for which the two action differences in the
sine functions are almost identical, resulting in a
non-oscillatory contribution to the trajectory sum. However, this
alone does not ensure a finite contribution to $\langle I^2 \rangle$:
In addition,
the ensemble averages of the winding numbers should not vanish. This
means that the trajectory pairs $\gamma_1$ and $\gamma_3$, and
$\gamma_2$ and $\gamma_4$ should be correlated for (at least) part of
their duration. However, since these trajectories exit the dot in
different contacts, they can not be identical.

\begin{figure}[t]
\epsfxsize=0.8\hsize
\hspace{0.01\hsize}
\epsffile{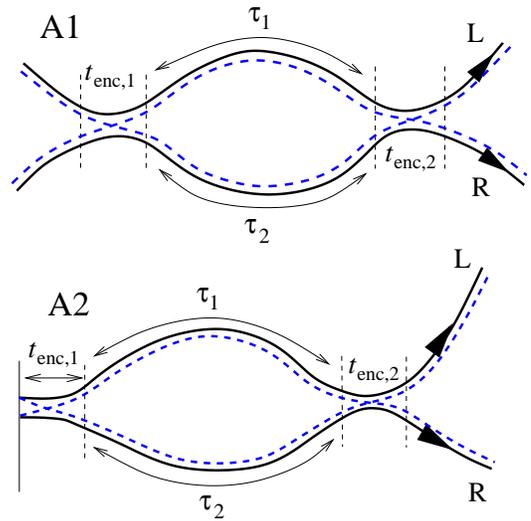}
\caption{(Color online)\label{fig:typea2} Trajectories of type A without and with
  correlated entry into the quantum dot (top and bottom, respectively).
\label{corr2}}
\end{figure}

The configurations of classical trajectories that meet these
requirements are the same as the trajectories that contribute to
conductance fluctuations.\cite{kn:brouwer2006} They are shown in Fig.\
\ref{typea}. The first type (labeled ``A'') consists of two pairs
of trajectories that undergo two separate small-angle
encounters. Between the encounters and the contacts $\gamma_1$ and
$\gamma_2$, and $\gamma_3$ and $\gamma_4$ are pairwise equal (up to
quantum uncertainties). Between the two encounters, $\gamma_1$ is paired
with $\gamma_3$ or $\gamma_4$, while $\gamma_2$ is paired with the
remaining trajectory, $\gamma_4$ or $\gamma_3$. The second type of
trajectories (labeled ``B'')
contains a closed loop, which is traversed one (extra)
time by two of the four trajectories.
Below we discuss both contributions separately. Since the calculations
closely follow the calculations of the conductance fluctuations, we
refer to Ref.\ \onlinecite{kn:brouwer2006} for details of the
formalism, and here restrict ourselves to those parts of the
calculation where we differ from Ref.\ \onlinecite{kn:brouwer2006}.

\subsection{contributions of type A}

Within the configurations of type A, one distinguishes
trajectories for which the first small-angle encounter fully resides
inside the quantum dot, such that the entrance of the pairs
trajectories $\gamma_1$ and $\gamma_2$, and $\gamma_3$ and $\gamma_4$
is uncorrelated, and a configuration of trajectories in which the
first small-angle encounter touches the entrance contact, so that all four
trajectories enter the dot together. The two situations are shown in
Fig.\ \ref{fig:typea2}. Since $\gamma_1$ and $\gamma_2$ exit through
the left contact, whereas $\gamma_3$ and $\gamma_4$ exit through the
right contact, the second encounter must fully reside inside the
quantum dot in all cases. We refer to the two cases shown in Fig.\
\ref{fig:typea2} as ``A1'' and ``A2''.

The durations of the encounters, defined as the time during which the
phase space distance between the trajectories is less than a
classical cut-off scale $c$, are denoted $t_{{\rm enc},1}$ and
$t_{{\rm enc},2}$, as depicted in Fig.\ \ref{fig:typea2}.
The durations of the uncorrelated
stretches between the encounters are denoted by $\tau_1$ and
$\tau_2$. We then expand
\begin{widetext}
\begin{eqnarray}
  \label{sinprod}
  4 \sin [( {{\cal S}_{\gamma_2}-{\cal S}_{\gamma_1}})/{\hbar}] \sin
  [({{\cal S}_{\gamma_4}-{\cal S}_{\gamma_3}})/{\hbar}] =
  e^{i ({\cal S}_{\gamma_2} - {\cal S}_{\gamma_1} +
  {\cal S}_{\gamma_3}-{\cal S}_{\gamma_4})/\hbar} -  e^{i ({\cal S}_{\gamma_2} -
  {\cal S}_{\gamma_1} + {\cal S}_{\gamma_4}-{\cal S}_{\gamma_3})/\hbar}
  + \mbox{c.c.}
\end{eqnarray}
and find quadruples of trajectories $\gamma_1$, $\gamma_2$,
$\gamma_3$, and $\gamma_4$ for which the exponents in Eq.\
(\ref{sinprod}) are small. Between the encounter regions and the
contacts, $\gamma_1$ and $\gamma_2$, and $\gamma_3$ and $\gamma_4$
are paired, resulting in only a small contribution to the action
differences in Eq.\ (\ref{sinprod}). In order to ensure that the
contribution from the stretch between the encounters is small as
well, one pairs $\gamma_1$ with $\gamma_3$ and $\gamma_2$ with
$\gamma_4$ between the encounters for the first term in Eq.\
(\ref{sinprod}), and $\gamma_1$ with $\gamma_4$ and $\gamma_2$ with
$\gamma_3$ for the second term. In the former case $\left<
W_1(\gamma_1) W_1(\gamma_3) \right> = C_1 (t_{{\rm enc},1} + t_{{\rm
enc},2} + \tau_1)$ and $\left< W_2(\gamma_2) W_2(\gamma_4) \right> =
C_2 (t_{{\rm enc},1} + t_{{\rm enc},2} + \tau_2)$. In the latter
case, one has $\left< W_1(\gamma_1) W_1(\gamma_3) \right> = C_1
(t_{{\rm enc},1} + t_{{\rm enc},2})$ and $\left< W_2(\gamma_2)
W_2(\gamma_4) \right> = C_2 (t_{{\rm enc},1} + t_{{\rm enc},2})$. At
each of the encounters we take a Poincar\'e surface of section. The
action differences $\Delta {\cal S} = {\cal S}_{\gamma_2} - {\cal
S}_{\gamma_1} + {\cal S}_{\gamma_3}-{\cal S}_{\gamma_4}$ or $\Delta
{\cal S} = {\cal S}_{\gamma_2} - {\cal S}_{\gamma_1} + {\cal
S}_{\gamma_3}-{\cal S}_{\gamma_4}$ that appear in the exponents in
Eq.\ (\ref{sinprod}) are then expressed in terms of the phase space
distances $s_j$ and $u_j$ between the two solid trajectories in
Fig.\ \ref{typea} along the stable and unstable directions in phase
space at each of the encounters,
  $\Delta {\cal S} = s_1 u_1 + s_2 u_2$.\cite{kn:spehner2003,kn:turek2003}
The encounter durations $t_{{\rm enc},1}$
and $t_{{\rm enc},2}$ are expressed in terms of
these coordinates as
\begin{equation}
  t_{{\rm enc},j} = \frac{1}{\lambda} \ln \frac{c^2}{|s_j u_j|},\ \
  j=1,2.
  \label{eq:tenc}
\end{equation}
Calculating the trajectory sum as in Refs.\
\onlinecite{kn:heusler2006,kn:braun2005,kn:brouwer2006,kn:brouwer2006c},
we then find
\begin{eqnarray}
\label{typea1}
\left< \Pi^L \Pi^R \right>_{A1} &=& \frac{8 \pi^2 N_1 N_2
  \tau_{\rm D}^2}{N^2}
\left( \prod_{j=1}^2 \int_0^\infty \frac{d \tau_j}{\tau_{\rm D}}
  e^{-\tau_j/\tau_{\rm D}} \right)
  \int_{-c}^c ds_1 du_1 ds_2 du_2
  \frac{e^{-i (s_1 u_1 + s_2 u_2)/\hbar - (t_{{\rm
  enc},1}+t_{{\rm enc},2})/\tau_{\rm D}}}{(2 \pi \hbar)^2
  t_{{\rm enc},1} t_{{\rm enc},2}}
  \nonumber \\ && \times
  C_1 C_2
  \left[ (t_{{\rm enc},1} +t_{{\rm enc},2}) (\tau_1 +\tau_2) + \tau_1 \tau_2 \right].
\end{eqnarray}
The factors $t_{{\rm enc},1}$
and $t_{{\rm enc},2}$ in the denominator in Eq.\ (\ref{typea1}) cancel
a spurious contribution from the freedom to take the Poincar\'e
surface of section at an arbitrary point in the encounter region.

We then make the variable change $u_j = c/\sigma_j$, $s_j = c x_j
\sigma_j$. In terms of the new variables, one has $t_{{\rm enc},j} =
\lambda^{-1} \ln (1/|x_j|)$.
Performing the integrals over $\tau_1$, $\tau_2$, $\sigma_1$, and
$\sigma_2$, one then finds
\begin{eqnarray}
\label{lasta1}
  \left< \Pi^L \Pi^R \right>_{A1} &=& \frac{32 N_1 N_2}{N^2} C_1 C_2
  \prod_{j=1}^2 \left[ r \tau_{\rm D} \lambda \int_0^1 dx_j
  x_j^\frac{1}{\lambda \tau_{\rm D}} \cos ( r x_j ) \right]
  \left[ 2 \tau_{\rm D} (t_{{\rm enc},1} +t_{{\rm enc},2}) + \tau_{\rm D}^2 \right],
\end{eqnarray}
where we abbreviated $r = c^2/\hbar$.

We now turn to compute contributions of type A2. Because the first
encounter touches the leads, the encounter duration $t_{{\rm enc},1}$
is no longer determined by the phase space coordinates $s_1$ and
$u_1$, but becomes an integration variable itself. The integration
range for $t_{{\rm enc},1}$ is $\lambda^{-1} \ln(c/|u_1|) < t_{{\rm enc},1} <
\lambda^{-1} \ln(c^2/|u_1 s_1|)$,\cite{kn:brouwer2006c} so that
\begin{eqnarray}
\label{firsta2}
\left< \Pi^L \Pi^R \right>_{A2} &=& \frac{8 \pi^2 N_1 N_2 \tau_{\rm
    D}^2}{N^2}
\left( \prod_{j=1}^2 \int_0^\infty \frac{d \tau_j}{\tau_{\rm D}}
e^{-\tau_j/\tau_{\rm D}} \right)  \int_{-c}^c ds_1 du_1 ds_2 du_2
  \int_{\lambda^{-1} \ln c/|u_1|}^{\lambda^{-1} \ln c^2/|s_1 u_1| }
    \frac{d t_{{\rm enc},1}}{\tau_D}
  \nonumber \\ && \times
  \frac{e^{-i (s_1 u_1 + s_2 u_2)/\hbar - (t_{{\rm
  enc},1}+t_{{\rm enc},2})/\tau_{\rm D}}}{(2 \pi \hbar)^2
  t_{{\rm enc},1} t_{{\rm enc},2}}
  C_1 C_2 \left[ (t_{{\rm enc},1} +t_{{\rm enc},2}) (\tau_1 +\tau_2) + \tau_1 \tau_2
    \right].
\end{eqnarray}
The second encounter time is still given by Eq.\ (\ref{eq:tenc})
above. As before, we
perform the variable change $u_j=c/\sigma_j$, $s_j=c x_j \sigma_j$,
$j=1,2$. Upon integration over $\tau_1$, $\tau_2$,
$\sigma_1$, $\sigma_2$, and $t_{{\rm enc},1}$ we then find
\begin{eqnarray}
\left< \Pi^L \Pi^R \right>_{A2} &=& - \frac{32 N_1 N_2}{N^2} C_1 C_2
  \left[ r \tau_{\rm D} \lambda \int_0^1 dx_2 x_2^{\frac{1}{\lambda
  \tau_{\rm D}}} \cos ( r x_2 ) \right]
  \nonumber \\ && \mbox{} \times
  r \tau_{\rm D} \lambda \int_0^1 dx_1 \cos ( r x_1 )
  \left[ x_1^{1/\lambda \tau_{\rm D}}
  (2 \tau_{\rm D} t_{{\rm enc},1} + 2 \tau_{\rm D} t_{{\rm enc},2}
  + 3 \tau_{\rm D}^2)
  - (2 \tau_{\rm D} t_{{\rm enc},2} + 3 \tau_{\rm D}^2) \right].
  \label{eq:IRLA2}
\end{eqnarray}
Here we reverted to the notation $t_{{\rm enc},1} =   \lambda^{-1} \ln
(1/|x_1|)$ in order to make contact with the result for $\langle \Pi^L
\Pi^R \rangle_{A1}$ obtained previously. The second term between the
square brackets $[\ldots]$ in the second line of Eq.\ (\ref{eq:IRLA2})
leads to a rapidly oscillating function of $r$ and will be neglected.
Combining the remaining terms with Eq.\ (\ref{lasta1}) and taking the
limit $\hbar \to 0$ ($r \to \infty$) while keeping the ratio
$\tau_{\rm E}/\tau_{\rm D}$ fixed, we find
\begin{eqnarray}
\left< \Pi^L \Pi^R \right>_{A} &=& -64  \frac{N_1 N_2}{N^2}
 C_1 C_2 \tau_{\rm D}^4 \left[ r \lambda \int_0^1 dx x^{{1}/{\lambda
 \tau_{\rm D}}} \cos (r x)\right]^2 \nonumber \\ &=&
 - 16 \pi^2 \frac{N_1N_2}{N^2} C_1 C_2 \tau_{\rm D}^2
 e^{-2 \frac{\tau_{\rm E}}{\tau_{\rm D}}},
  \label{eq:PiLRAfinal}
\end{eqnarray}
\end{widetext}
where the Ehrenfest time is defined as
\begin{equation}
\label{deftaue}
  \tau_{\rm E} = \frac{1}{\lambda} \ln r =
  \frac{1}{\lambda} \ln \frac{c^2}{\hbar}.
\end{equation}
Note that the limit $\hbar \to 0$ at fixed $\tau_{\rm E}/\tau_{\rm
D}$, taken above, can be realized by narrowing the lead openings
such that the dwell time increases. While this procedure may change
the trajectories contributing to pumping, their statistical
properties should be (almost) unaffected. In particular, the
Lyapunov exponent will converge to its value in the closed cavity.

\subsection{contributions of type B}

The trajectory configurations of type B provide the dominant
contributions to the pumped current at large Ehrenfest times. Each
pair $(\gamma_1,\gamma_2)$ and $(\gamma_3,\gamma_4)$ consists of a
`short' trajectory and a `long' trajectory, where the long trajectory
differs from the short trajectory by winding once around a
periodic orbit, see Fig.\ \ref{typea}. (Strictly speaking, one should
say that the long trajectory winds one extra time around the periodic
orbit, because all trajectories involved will wind multiple times
around the periodic orbit if $\tau_{\rm E}$ is much longer than the
period $\tau_{p}$ of the periodic orbit.) In the trajectory sum
(\ref{eq:Isemi}), each trajectory can be the short one. Following
Ref.\ \onlinecite{kn:brouwer2006}, we parametrize these configurations
by Poincar\'e surfaces of sections which measure the phase space
coordinates $(s_1,u_1)$ and $(s_2,u_2$) of two
short trajectories with respect to
the stable and unstable manifolds of the periodic orbit.

\begin{figure}[t]
\epsfxsize=0.9\hsize
\hspace{0.01\hsize}
\epsffile{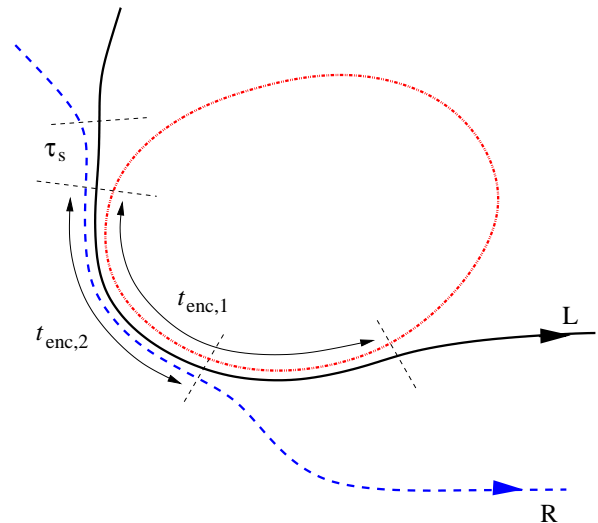}
\caption{(Color online)\label{deft} Definitions of the relevant durations for
  trajectories of type B. The figure shows the short trajectories from
  each pair, as well as the periodic orbit. The long trajectories from
  each pair, which are not shown, differ from the short trajectories
  by one revolution around the periodic orbit. The figure shows the
  durations $t_{{\rm enc},1}$ and $t_{{\rm enc},2}$ of the encounters
  of the short trajectories with the periodic orbit, as well as the
  time $\tau_s$ that the trajectories are correlated before arriving
  at the periodic orbit. A similar time $\tau_u$ measures the time the
  trajectories are correlated after departure from the periodic
  orbit. In the example shown in the figure, the trajectories depart
  at different points, so that $\tau_u = 0$.}
\end{figure}

It will be useful to define several relevant times. The two short
trajectories in the configuration are depicted in Fig. \ref{deft}.
The time for which the short trajectories are correlated
with the periodic trajectory (phase space distances smaller than the
cutoff $c$)
is denoted $t_{{\rm enc},1}$ [for the short trajectory of the pair
$(\gamma_1,\gamma_2)$] and $t_{{\rm enc},2}$ [for the short
trajectory of the pair $(\gamma_3,\gamma_4)$].
By our construction, the longer trajectory
in the pair (not drawn in Fig. \ref{deft}) will be correlated
with the periodic trajectory for a time $t_{{\rm enc},j}+\tau_p$, $j=1,2$,
where $\tau_p$ is the period of the periodic orbit.
Since the trajectories may leave the neighborhood of the periodic
trajectory together, the correlations may extend away from
the periodic trajectory. If applicable, we denote the time for which
the trajectories are correlated with each other before they arrive to
the periodic orbit by $\tau_s$. Similarly, $\tau_u$ denotes the
eventual correlation time after the trajectories leave the periodic
orbit. If $\tau_s \neq 0$, the encounter region extends away from the
periodic orbit and may touch the entrance contact. Hence, we
distinguish contributions of type B1 and type B2, where type B1 refers
to those trajectories for which the encounters fully reside in the
quantum dot, and type B2 refers to configurations of trajectories with
correlated entry (encounter that touches the entrance lead opening).
Since the trajectory pairs $(\gamma_1,\gamma_2)$
and $(\gamma_3,\gamma_4)$ exit through different contacts, one does
not have to consider the possibility that the encounter region
touches the exit lead opening.
The Poincar\'e
surfaces of section are taken at points where the short trajectories
are correlated with the periodic orbit. The distance
between the two Poincar\'e surfaces of section is measured by the
travel time $t_{12}$ along the periodic orbit, where we require $|t_{12}| <
\tau_{p}/2$.

Even the shortest trajectory in a pair can wind around
the periodic trajectory several times. Thus, it is
possible that $t_{{\rm enc},j}>\tau_p$. In order to separate out full
revolutions, we write
\begin{equation}
   t_{{\rm enc},j} = n_j \tau_p+\tilde{t}_{{\rm enc},j},
\end{equation}
where $n_j$ is a non-negative integer and
$0 \le \tilde{t}_{{\rm enc},j} < \tau_p$.
The use of $\tilde{t}_{{\rm enc},j}$ complicates some of the calculations
since this time is not a continuous function of the surface of section
coordinates. On the other hand, this time appears naturally due to the
random Gaussian character of the winding numbers $W_{1,2}$.
\begin{figure}[t]
\epsfxsize=0.9\hsize
\hspace{0.01\hsize}
\epsffile{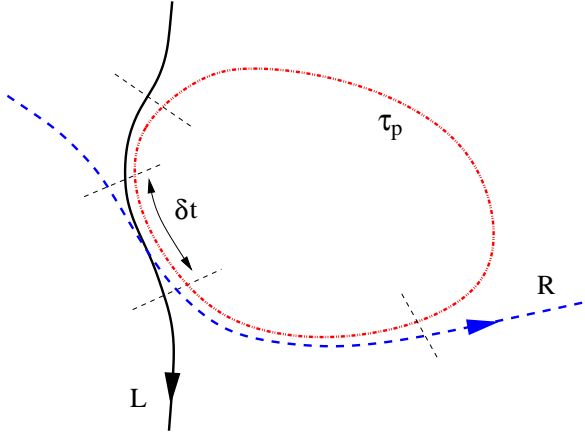}
\caption{(Color online)\label{dt} The overlap time is defined as the overlap of the
  two encounter times $t_{{\rm enc},1}$ and $t_{{\rm enc},2}$.}
\end{figure}

Without loss of generality (but with inclusion of a combinatorial factor
two), we take $\gamma_1$ to be one of the short
trajectories. Expanding the product of sines as in
Eq.\ (\ref{sinprod}), we then see that smallness of the the action
difference requires $\gamma_3$ to be the other short trajectory for
the first term in Eq.\ (\ref{sinprod}), whereas $\gamma_4$ is the
other short trajectory in the second term in Eq.\ (\ref{sinprod}). In
the former case, one has
\begin{widetext}
\begin{eqnarray*}
  \left< W_1(\gamma_1) W_1 (\gamma_3) \right> &=& C_1 [ n_1 n_2 \tau_p
  + n_1 \tilde{t}_{{\rm enc},2} +n_2 \tilde{t}_{{\rm enc},1} + \delta \tilde{t} +
  \tau_s + \tau_u ] \\
  \left< W_2(\gamma_2) W_2 (\gamma_4) \right> &=& C_2 [ (n_1+1)
  (n_2+1) \tau_p
  + (n_1+1) \tilde{t}_{{\rm enc},2} + (n_2+1) \tilde{t}_{{\rm enc},1} + \delta
  \tilde{t} + \tau_s + \tau_u ].
\end{eqnarray*}
The time $\delta \tilde{t}$ is the overlap
of the `remainders' $\tilde t_{{\rm enc},1}$ and $\tilde t_{{\rm enc},2}$, see
Fig.~\ref{dt}. In the latter case, one has
\begin{eqnarray*}
  \left< W_1(\gamma_1) W_1 (\gamma_3) \right> &=& C_1 [ n_1 (n_2+1) \tau_p
  + n_1 \tilde{t}_{{\rm enc},2} +(n_2+1) \tilde{t}_{{\rm enc},1} + \delta \tilde{t} +
  \tau_s + \tau_u ] \\
  \left< W_2(\gamma_2) W_2 (\gamma_4) \right> &=& C_2 [ (n_1+1)
  n_2 \tau_p
  + (n_1+1) \tilde{t}_{{\rm enc},2} + n_2 \tilde{t}_{{\rm enc},1} + \delta
  \tilde{t} + \tau_s + \tau_u ].
\end{eqnarray*}
The resulting contribution of type B1 then reads
\begin{eqnarray}
\label{typeb1}
\left< \Pi^L \Pi^R \right>_{B1} &=& 16 \pi^2 \frac{N_1 N_2\tau_{\rm
    D}}{N^2}
  C_1 C_2
  \int \frac{d \tau_p}{\tau_{\rm D}} e^{-{\tau_p}/{\tau_{\rm D}}}
  \int_{-\tau_p/2}^{\tau_p/2} dt_{12}  \int_{-c}^{c} ds_1 du_1 ds_2 du_2
  \frac{e^{i{(u_1 s_1 - u_2 s_2)}/{\hbar}-({\tau_s+\tau_u})/{\tau_{\rm
    D}}}}{(2 \pi \hbar)^2 t_{{\rm enc},1} t_{{\rm enc},2}} \nonumber \\ &&
  \mbox{} \times
  \left[ \tau_p \left( \delta \tilde{t} + \tau_s + \tau_u \right) -
  \tilde{t}_{{\rm enc},1} \tilde{t}_{{\rm enc},2} \right].
\end{eqnarray}
The contributions of type B2 are calculated in a way very similar to
the one used for type A2,
\begin{eqnarray}
\label{typeb2}
\left< \Pi^L \Pi^R \right>_{B2} &=& 16 \pi^2
  \frac{N_1 N_2 \tau_{\rm D}}{N^2} C_1 C_2
\int \frac{d \tau_p}{\tau_{\rm D}} e^{-{\tau_p}/{\tau_{\rm D}}}
\int_{-\tau_p/2}^{\tau_p/2} dt_{12} \int_{-c}^{c} ds_1 du_1 ds_2 du_2
\frac{e^{i{(u_1 s_1 - u_2 s_2)}/{\hbar}-{\tau_u}/{\tau_{\rm D}}}}{(2 \pi
  \hbar)^2 t_{{\rm enc},1} t_{{\rm enc},2}} \nonumber \\ && \mbox{} \times
 \left\{ \left( 1 -  e^{-{\tau_s}/{\tau_{\rm D}}} \right) \left[ \tau_p \left( \delta \tilde{t} + \tau_{\rm D} + \tau_u \right) - \tilde{t}_{{\rm enc},1} \tilde{t}_{{\rm enc},2} \right] -  e^{-\frac{\tau_s}{\tau_{\rm D}}} \tau_p \tau_s\right\}.
\end{eqnarray}
Adding both contributions, we then obtain
\begin{eqnarray}
\label{typeb}
  \left< \Pi^L \Pi^R \right>_{B} &=&
  16 \pi^2 \frac{N_1 N_2 \tau_{\rm D}}{N^2} C_1 C_2
\int \frac{d \tau_p}{\tau_{\rm D}} e^{-\frac{\tau_p}{\tau_{\rm D}}}
  \int_{-\tau_p/2}^{\tau_p/2} dt_{12}
  \int_{-c}^{c} ds_1 du_1 ds_2 du_2
  \frac{e^{i{(u_1 s_1 - u_2 s_2)}/{\hbar}-{\tau_u}/{\tau_{\rm D}}}}{(2 \pi
  \hbar)^2 t_{{\rm enc},1} t_{{\rm enc},2}} \nonumber \\ && \mbox{} \times
\left\{ \left(  1 -  e^{-{\tau_s}/{\tau_{\rm D}}} \right) \tau_p \tau_{\rm D} + \tau_p (\delta \tilde{t} +\tau_u) - \tilde{t}_{{\rm enc},1} \tilde{t}_{{\rm enc},2}\right\}.
\end{eqnarray}
\end{widetext}

In App.\ \ref{apptimes} it is shown that the contribution from the
combination $\tau_p \delta \tilde{t} - \tilde{t}_{{\rm enc},1} \tilde{t}_{{\rm enc},2}$
between the curly brackets $\{ \ldots \}$ is of order $C_1 C_2
\tau_{\rm D}/\lambda$, which can be neglected with respect to $\langle
\Pi^L \Pi^R \rangle_A$.
The contribution from the term $\tau_p \tau_u$ is a rapidly
oscillating function of $r = c^2/\hbar$ and is omitted. The only
nonzero contribution arises from the term $\tau_p \tau_{\rm D}$, which
is the only term that is multiplied by both $\exp(-\tau_s/\tau_{\rm
  D})$ and $\exp(-\tau_u/\tau_{\rm D})$. This factor $\tau_p \tau_{\rm
  D}$ can be taken out of the integral over the phase space variables.
The resulting integral was computed in Ref.\
\onlinecite{kn:brouwer2006}, with the result
\begin{eqnarray}
\label{tempb}
  \left< \Pi^L \Pi^R \right>_{B} &=&
  - 8 \pi^2 \frac{N_1N_2}{N^2} C_1 C_2 \int d \tau_p
  \tau_p e^{-{\tau_p}/{\tau_{\rm D}}}
 \nonumber \\ && \mbox{} \times (1 - e^{-2 \tau_{\rm
    E}/\tau_{\rm D}}).
\end{eqnarray}
In Eq. (\ref{tempb}) we omitted terms proportional to $e^{-\lambda
\tau_p}$. Typical periodic trajectories contributing to Eq.
(\ref{tempb}) have duration $\sim \tau_{\rm D}$. In the limiting
procedure discussed below Eq. (\ref{deftaue}) these contributions
scale as $\exp(-\lambda \tau_{\rm D}) \to 0$, justifying this
approximation.
The integral over the period of the periodic trajectory is easily calculated,
resulting in
\begin{equation}
\label{tempb2}
\left< \Pi^L \Pi^R \right>_{B}= - 8 \pi^2 \frac{N_1N_2}{N^2} C_1 C_2 \tau_{\rm D}^2
\left( 1- e^{-2 {\tau_{\rm E}}/{\tau_{\rm D}}}\right).
\end{equation}


Combining both contributions together, we find that the total variance
of the pumped current is
\begin{equation}
\label{res}
\left< I^2 \right> = 2 \frac{N_1N_2}{N^2} {\cal A}^2 e^2 \omega^2 C_1 C_2 \tau_{\rm D}^2
\left( 1+ e^{-2\frac{\tau_{\rm E}}{\tau_{\rm D}}}\right).
\end{equation}
Equation (\ref{res}) agrees with random matrix theory in the limit
$\tau_{\rm E} \rightarrow
0$.\cite{kn:brouwer1998,kn:shutenko2000,kn:vavilov2001a}
At finite $\tau_{\rm E}$, the variance
of the pumped current is reduced below the random matrix value, but
the reduction is not more than a factor two.

When $\tau_{\rm E} \ll \tau_{\rm D}$ the pumped current is dominated
by trajectories of type A. For $\tau_{\rm E} \gg \tau_{\rm D}$
the pumped current is
carried by trajectories of type B. Since such trajectory configurations
involve a closed loop they are associated with fluctuations of the
density of states. Although this scenario is very similar to that of
the Ehrenfest-time dependence of the conductance fluctuations, the
pumped current has a $\tau_{\rm E}$-dependent part, whereas the
conductance fluctuations in a chaotic quantum dot are fully $\tau_{\rm
E}$ independent. The difference occurs, because the mean total
duration of the trajectories involved in the internal loop for
trajectories of type A is twice the mean duration of the internal loop
for trajectories of type B. Conductance fluctuations are insensitive
to the loop duration (as long as the trajectories stay inside the
quantum dot), but the pumped current is not.

Equation (\ref{res}) is valid for systems with and without
time reversal symmetry. This is well known in terms of random
matrix theory.\cite{kn:brouwer1998,kn:shutenko2000} We have verified
this explicitly in the semiclassical approach, see App.\ \ref{TRS}.
We also verified that
calculation of the current variance using the correlator $\langle
(\Pi^L)^2 \rangle$, while technically more involved, gives the same
result for the pumped current, as required by unitarity.

\section{Numerical simulation}
\label{numerics}

Despite the remarkable similarity of the semiclassical calculation of
the pumped current and the semiclassical calculation of the
conductance fluctuations, the former shows a dependence on the
Ehrenfest time, whereas the latter does not. In this section we report
numerical simulations of the pumped current and compare these
with our theoretical predictions.

Because the computational cost of numerical simulation of
two-dimensional quantum dots with large Ehrenfest times is
prohibitive, Jacqoud, Schomerus, and Beenakker
suggested to simulate one dimensional chaotic maps
instead.\cite{kn:jacquod2003} The
phenomenology of chaotic maps is identical to that of chaotic
cavities,\cite{kn:fishman1982,kn:izrailev1990}
but the computational cost of simulating a map is
significantly lower than that of simulating a cavity.  Chaotic maps
have been successfully used to study the Ehrenfest time dependence of a
range of properties of chaotic quantum dots.\cite{kn:jacquod2003,kn:tworzydlo2003,kn:jacquod2004,kn:tworzydlo2004b,kn:tworzydlo2004c,kn:rahav2005,kn:rahav2006,kn:schomerus2005}

The quantum map propagates a finite state vector of size $M=1/2 \pi
\hbar$ in time,
\begin{equation}
\psi (t+1) = {\cal F} \psi (t),
\end{equation}
where ${\cal F}$ is the Floquet operator of the map. Assigning two
consecutive sets of $N_1$ and $N_2$ elements of the vector $\psi$ to
contacts, the map can be used to construct a $(N_1+N_2)$-dimensional
scattering matrix $S$
according to the rule\cite{kn:fyodorov2000,kn:jacquod2003}
\begin{equation}
  S(\varepsilon) = {\cal P} \left[ 1 - e^{i \varepsilon} {\cal F} {\cal Q} \right]^{-1}
e^{i \varepsilon} {\cal F} {\cal P}^T
\end{equation}
where ${\cal P}$ is a $(N_1+N_2)\times M$ matrix projecting on the
lead sites, ${\cal Q}= 1 -{\cal P}^T{\cal P}$, and $\varepsilon$ is
the quasi energy.

The map used in our simulations is the
'three-kick quantum rotator',~\cite{kn:tworzydlo2004b}
which is specified by the Floquet operator,
\begin{equation}
\label{3kickf}
{\cal F}_{mn} = \left( X U Y^* U Y U X \right)_{mn}
\end{equation}
where
\begin{eqnarray}
 Y_{mn} & = & \delta_{mn} e^{i (\gamma M /6 \pi) \cos (2 \pi m/M)}, \nonumber \\
 X_{mn} & = & \delta_{mn} e^{-i (M/12 \pi) V (2 \pi m /M)}, \nonumber \\
 U_{mn} & = & M^{-1/2} e^{-i \pi/4} e^{i \pi (m-n)^2 /M}.
\end{eqnarray}
In this model $M$ is even, but not a multiple of $3$. The kick potential is
given by
\begin{equation}
 V(\theta) = K \cos (\pi q/2) \cos \theta + \frac{1}{2} K \sin (\pi q/2) \sin 2 \theta,
\end{equation}
where $q$ breaks the parity symmetry of the model.~\cite{kn:bluemel1992b}
The parameter $\gamma$ plays the role of a magnetic field resulting in the
breaking of time reversal symmetry. This model was used in
Refs. \onlinecite{kn:tworzydlo2004b} and \onlinecite{kn:rahav2006}
to compute the weak localization correction of the conductance.

There are two parameters in this model: The chaoticity parameter $K$
and the 'magnetic field' $\gamma$. We will use $K$ and
$\gamma$ as the time-dependent parameters that drive the quantum pump.
Although both $K$ and $\gamma$ appear in the classical map and
thus affect classical trajectories, the
variations used in our simulations are sufficiently small that the
changes of the classical dynamics can be neglected
within the mean dwell time $\tau_{\rm D} = M/(N_1+N_2)$.

In the simulations we took equal channel numbers in both contacts,
$N_1=N_2=N/2$. The semiclassical limit is taken by increasing the
dimension $M=1/2 \pi \hbar$ of the system, while keeping the dwell
time $\tau_{\rm D} = M/N$ fixed. This means that $N$ is scaled
proportional to $M$. This way we are certain that there are no
variations in the classical dynamics while decreasing $\hbar$.
With these definitions the Ehrenfest time is estimated to be
\begin{equation}
\tau_{\rm E} \simeq \frac{1}{\lambda} \ln N,
  \label{eq:taunum}
\end{equation}
plus an $N$-independent constant. The Lyapunov exponent $\lambda$ is
calculated independently from a simulation of the classical
counterpart of the map.

The pumping strength depends on the two parameters $C_K$ and
$C_{\gamma}$ that describe the rate at which variations of $K$ and
$\gamma$ affect transport. (In the semiclassical theory, these
parameters were called $C_1$ and $C_2$.) In order to calculate
$C_K$ and $C_{\gamma}$, we first performed separate numerical
simulations of the quantities
\begin{eqnarray}
  y_{K} &=& \sum_{\alpha=1}^{N_1} \sum_{\beta=N_1+1}^{N}
  \left| \frac{\partial S_{\alpha\beta}}{\partial K} \right|^2,
  \nonumber \\
  y_{\gamma} &=& \sum_{\alpha=1}^{N_1} \sum_{\beta=N_1+1}^{N}
  \left| \frac{\partial S_{\alpha\beta}}{\partial \gamma} \right|^2,
\end{eqnarray}
and compared these to the result of a semiclassical calculation,
\begin{equation}
  y_{K,\gamma} = \frac{N_1 N_2}{N} C_{K,\gamma} \tau_{\rm D}.
\end{equation}
We then calculated the kernels $\Pi^L$ and $\Pi^R$ numerically, and
considered the ratio
\begin{equation}
  R = \frac{-N_1 N_2}{8 \pi^2} \left< \frac{\Pi^L \Pi^R}{y_K y_{\gamma}}\right>.
\end{equation}
The prediction of the semiclassical theory is
\begin{equation}
  R = 1 + e^{-2 \tau_{\rm E}/\tau_{\rm D}}.
\end{equation}
\begin{figure}[t]
\epsfxsize=0.9\hsize
\hspace{0.01\hsize}
\epsffile{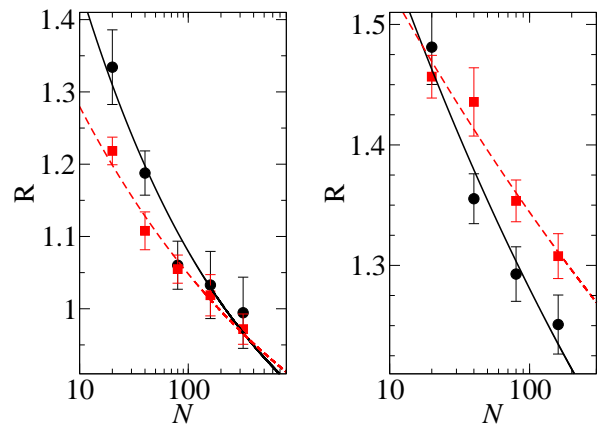}
\caption{(Color online)\label{num} The normalized current variance $R$, as a function
of $N$. The symbols denote simulation results for the three-kick model
with $K=10$, $q=0.2$ and $\gamma=5$ (circles) or $\gamma=10$ (squares).
Results for $\tau_{\rm D}=5$ (left panel) and $\tau_{\rm D}=10$ (right panel)
are shown.
The solid and dashed line are fits to theory, as described in text.}
\end{figure}

The normalized pumping current variance $R$ is ensemble averaged
by variation of the quasi energy $\varepsilon$ and lead positions.
Since the variation of the lead positions may result in large classical
variations, we use
the same set of lead positions for different values of $M$.
The factors $\Pi^L \Pi^R$, $y_K$ and $y_{\gamma}$ are first averaged
over the quasienergy. This average can be done analytically.~\cite{kn:rahav2006}
Then $R$ is averaged over $2\, 000$ different lead positions.

We have checked that when the dwell time is increased the pumping current variance
tends to the random matrix limit $R=2$. (Data not shown.  The trend can be
seen by comparing the left panel of Fig.~\ref{num}  [where $\tau_{\rm D}=5$]
to the right panel [where $\tau_{\rm D}=10$].)
However, in
order to access the regime in which $\tau_{\rm E}$ and $\tau_{\rm D}$
are comparable, we have to take the rather small value of $\tau_{\rm
  D} = 5$. In this case, full ergodicity is not achieved during the
time $\tau_{\rm D}$, which results in a deviation from the random
matrix limit even for small Ehrenfest times. A similar deviation from
random matrix theory was observed in numerical simulations of weak
localization and universal conductance
fluctuations.\cite{kn:tworzydlo2004,kn:tworzydlo2004b,kn:tworzydlo2004c,kn:jacquod2004,kn:rahav2005,kn:rahav2006,kn:jacquod2006}

While the necessity to consider short dwell times prevents a
quantitative comparison of simulations and
theory, our simulations, which are shown in Fig.\ \ref{num}, show an
unambiguous dependence on $\tau_{\rm E}$. To quantify the dependence
we have fitted the simulation results to a function of the form
$R=A+B \exp[-(2/\lambda \tau_{\rm D})\ln N]$.
The values of the Lyapunov exponent are obtained from a simulation
of the classical map~\cite{kn:tworzydlo2004b}
corresponding to (\ref{3kickf}). We find $\lambda \simeq 1.43$
for $\gamma=5$ and $\lambda \simeq 2.04$ for $\gamma=10$.
The fit constants are found to be $A \simeq 0.74$ and $B \simeq 1.18$
for $\gamma=5$ and $\tau_{\rm D}=5$;  $A \simeq 0.64$ and $B \simeq 1.0$
for $\gamma=10$ and $\tau_{\rm D}=5$ ;$A \simeq 0.56$ and $B \simeq 1.37$
for $\gamma=5$ and $\tau_{\rm D}=10$; and $A \simeq 0.61$ and $B \simeq 1.15$
for $\gamma=10$ and $\tau_{\rm D}=10$.
 Deviations from the theoretical values $A = B = 1$ are
attributed to the nonuniversal corrections (caused by the small dwell
time $\tau_{\rm D}$). The unknown additive constant in the
expression (\ref{eq:taunum}) for the Ehrenfest time is absorbed in the
parameter $B$ and may cause further deviations from the predicted
value $B=1$.

As mentioned above, the small dwell times used in the
simulation cause non-universal and systematic deviations from the
theoretical prediction.
Increasing the dwell time is numerically
expensive. It also strongly suppresses the dependence of $R$ on the number
of channels $N$. With these limitations, we conclude that, while the
simulation does not unambiguously confirm the semiclassical theory,
it certainly is not inconsistent with it: The data shows a clear
dependence on the Ehrenfest time, whereas the nonzero
fitted values for $A$ clearly
point to a finite value of $R$ in the classical
limit $N \rightarrow \infty$.

\section{Conclusion}
\label{disc}

In this article we presented a semiclassical theory of the current
pumped through a ballistic and chaotic quantum dot with two
time-dependent external perturbations. In a quantum pump, the external
perturbations couple to the phase of the electronic wavefunction,
not to the classical dynamics, so that the pumped current is generated
through quantum interference only. Our semiclassical theory identifies
the two classes of interfering trajectories that contribute to the
root-mean-square current $\mbox{rms}\, I = \langle I^2
\rangle^{1/2}$. These are shown in Fig.\ \ref{typea}. In each case,
the semiclassical contribution for $\langle I^2 \rangle$ originates
from two pairs of classical trajectories which meet at small-angle
encounters. The interference arises from time-dependent
phase differences accumulated between the encounters.

Our semiclassical calculation shows that the action of the external
perturbations during the small-angle encounters does not generate a
direct current. This follows, {\em e.g.}, from the final expressions
(\ref{eq:PiLRAfinal}) and (\ref{tempb}) for the two contributions to the
pumped current, which contain the mean times
of the interference loops but not the encounter times. Although such a
fact appears natural for a quantum interference effect --- nontrivial
interference can occur only if trajectories are separated in phase
space \mbox{---,} within the semiclassical formalism it follows only after
cancellation between contributions of trajectories with and without
correlated entry. In this respect, our calculation again emphasizes the
necessity to treat trajectory configurations with and without
correlated entry on the same
footing.\cite{kn:brouwer2006,kn:brouwer2006c,kn:whitney2006,kn:jacquod2006}

The main goal of our calculation was to provide a
quantitative description of the pumped
current if the Ehrenfest time $\tau_{\rm E}$ is comparable to the mean
dwell time $\tau_{\rm D}$ in the quantum dot.
The Ehrenfest time is the time required for the chaotic
classical dynamics to separate two phase space points initially a
Fermi wavelength apart to classical distance. As such, $\tau_{\rm E}$
presents a threshold time before which no quantum interference can
occur. While we do find that the pumped current is sensitive to the
ratio $\tau_{\rm E}/\tau_{\rm D}$, the pumped current remains nonzero,
even in the limit $\tau_{\rm E} \gg \tau_{\rm D}$. In the limit $\tau_{\rm
  E} \gg \tau_{\rm D}$, the trajectories that contribute to the pumped
current wind many times around a periodic orbit. The escape
probability for such a configuration depends on the period of the
periodic orbit, not on the total duration of the trajectories
involved. This explains why such trajectories continue to exist
even if their duration far exceeds the mean dwell time $\tau_{\rm D}$.

\acknowledgments

We would like to thank Maxim Vavilov for discussions.
This work was supported by the NSF under grant no.\ DMR 0334499 and
by the Packard Foundation.

\begin{appendix}

\section{Remainder times integration}
\label{apptimes}

In this appendix we show that the integral
\begin{eqnarray}
\label{intapp}
  J &=& \int d \tau_p e^{-{\tau_p}/{\tau_{\rm D}}}
\int_{-{\tau_p}/{2}}^{{\tau_p}/{2}} dt_{12} \int_{-c}^{c} ds_1
du_1 ds_2 du_2 \nonumber \\ && \mbox{} \times
\frac{e^{i(u_1 s_1 - u_2 s_2) /\hbar - \tau_{s}/\tau_{\rm D}}}{\hbar^2 t_{{\rm enc},1}
  t_{{\rm enc},2}} \left[ \tau_p \delta \tilde{t} - \tilde{t}_{{\rm enc},1}
  \tilde{t}_{{\rm enc},2} \right], ~~~
\end{eqnarray}
is of order $\tau_{\rm D}/\lambda$. After the replacement of
$\tau_s$ by $\tau_u$, which is allowed by symmetry, this proves the
claim that the terms proportional to $\tau_p \delta \tilde t - \tilde
t_{{\rm enc},1} \tilde t_{{\rm enc},2}$ in the curly brackets $\{
\ldots \}$ in Eq.\ (\ref{typeb}) does not contribute to the correlator
$\langle \Pi^L \Pi^R \rangle$.


First, we point out that the integral over $t_{12}$ vanishes if one
omits the factor $\exp[-{\tau_s}/{\tau_{\rm D}}]$ from the expression
for $J$. Hence, we may replace the factor
$\exp[-{\tau_s}/{\tau_{\rm D}}]$ in Eq.\ (\ref{intapp})
by $\exp[-{\tau_s}/{\tau_{\rm D}}] - 1$,
\begin{widetext}
\begin{eqnarray}
\label{intapp-2}
  J &=& \int d \tau_p e^{-{\tau_p}/{\tau_{\rm D}}}
  \int_{-{\tau_p}/{2}}^{{\tau_p}/{2}} dt_{12} \int_{-c}^{c} ds_1 du_1
  ds_2 du_2 \frac{e^{i(u_1 s_1 - u_2 s_2) /\hbar}}{\hbar^2 t_{{\rm
  enc},1} t_{{\rm enc},2}}
  \left( e^{-{\tau_s}/{\tau_{\rm D}}}-1\right) \left( \tau_p \delta
  \tilde{t} - \tilde{t}_{{\rm enc},1} \tilde{t}_{{\rm enc},2} \right).
\end{eqnarray}
The contributions for this integral are from trajectories $\gamma_1$,
$\gamma_2$, $\gamma_3$, and $\gamma_4$ with
correlated arrival at the periodic orbit,
since the integrand vanishes if $\tau_s = 0$.
On the other hand, the departure from the periodic orbit
need not be correlated.

Following Ref. \onlinecite{kn:brouwer2006} we perform a change of
variables using $s_j=c/\sigma_j$, $u_j=c x_j \sigma_j$, $j=1,2$, and
$t_{12}^\prime = t_{12} - \lambda^{-1} \ln \left( c/|s_1|\right)
+\lambda^{-1} \ln \left( c/|s_2|\right)$. This change of variables
amounts to choosing the Poincar\'e surfaces of section at the point
where each trajectory first comes within a phase space distance $c$
from the periodic orbit. The time $t_{12}^\prime$ is the distance
between the new Poincar\'e surfaces of
section, measured as the travel time along the periodic orbit. The
integrations over $\sigma_j$, $j=1,2$, can be done, and
cancel the factors $t_{{\rm enc},j}$ in the denominator.
The resulting expression for $J$ is
\begin{eqnarray}
  J &=& 2 \lambda^2 r^2 \int d \tau_p
  e^{-{\tau_p}/{\tau_{\rm D}}}
  \int_{-{\tau_p}/{2}}^{{\tau_p}/{2}} d t_{12}^\prime \sum_{\pm}
  \int_{-1}^1 dx_1 dx_2
  \left(  e^{-{\tau_s}/{\tau_{\rm D}}}-1\right) \cos \left( rx_1 \mp r
  x_2\right) \left( \tau_p \delta \tilde{t} -
  \tilde{t}_{{\rm enc},1} \tilde{t}_{{\rm enc},2} \right),
\end{eqnarray}
\end{widetext}
where the sum over the signs originates from the possible relative
signs between $s_1$ and $s_2$ and $r = c^2/\hbar$.
We perform another change of variables,\cite{kn:brouwer2006} defining
$u_1=x_1$, $u_2=\pm x_2$ and $w=\pm e^{\lambda t_{12}^\prime}$.
Both the integration over $t_{12}^\prime$ and the summation over the signs are combined
into the integral over $w$. A straightforward calculation leads to
\begin{eqnarray}
  \label{factorint}
  J &=& 2 \lambda r^2 \int d \tau_p
  e^{-\frac{\tau_p}{\tau_{\rm D}}}
  \int \frac{d w}{|w|} \int_{-1}^1 du_1 du_2
  \left(
  e^{-{\tau_s}/{\tau_{\rm D}}}-1\right)
  \nonumber \\ && \mbox{} \times
  \cos \left( r u_1 - r
  u_2\right) \left( \tau_p \delta \tilde{t} -
  \tilde{t}_{{\rm enc},1} \tilde{t}_{{\rm enc},2} \right).
\end{eqnarray}

In Eq.\ (\ref{factorint}), the time $\tau_s$ depends only on the
location where the trajectories in each pair become correlated with
the periodic orbit. Following Ref.\ \onlinecite{kn:brouwer2006}, we
set
\begin{equation}
  \tau_s = \frac{1}{\lambda} \ln \left[\frac{\max(w,1/w)-1}{b-1} \right],
\end{equation}
where $b$ is a numerical constant of order unity.
Because of the factor $\exp(-\tau_s/\tau_D) - 1$, only $|w|$
close to unity, corresponding to $\lambda |t_{12}'| \lesssim 1$,
contributes to the integral. This implies that the overlap time
$\delta \tilde t$ can be approximated by
\begin{equation}
  \delta \tilde t = \min(\tilde{t}_{{\rm enc},1},\tilde{t}_{{\rm enc},2} ),
\end{equation}
where
\begin{equation}
  \tilde{t}_{{\rm enc},j} =
  \frac{1}{\lambda} \ln(1/|u_j|)\ \mod \tau_p.
  \label{eq:mod}
\end{equation}
The terms neglected in this approximation can, at most, lead to
corrections of order $1/\lambda \tau_{\rm D}$.

The integrals over $w$ and $u_1$, $u_2$ factorize. Integrating
over $w$ gives a factor of order $1/\lambda \tau_{\rm D}$. The
remaining integral over $u_1$ and $u_2$ is complicated due to the
discontinuity of the encounter times when they reach the period
$\tau_p$, see Eq.\ (\ref{eq:mod}). We deal with this problem by
separating the integral into domains where the integrand is
continuous. The integral of the term proportional to $\tilde t_{{\rm
enc},1} \tilde t_{{\rm enc},2}$ factorizes to a product of two
integrals. They are given by
\begin{eqnarray}
J_{1,j} &=& r \int_0^1 du_j \cos (r u_j) \tilde{t}_{{\rm enc},j}
  \nonumber \\ &=&
  \label{tempj1}
  \frac{1}{\lambda} \int_0^1 \frac{d u_j}{u_j} \sin (r u_j) -
  \tau_p f(r),
\end{eqnarray}
where
\begin{equation}
  f(r) = \sum_{m=0}^\infty
  \sin \left( r e^{- (m+1) \lambda \tau_p} \right).
  \label{eq:fr}
\end{equation}
In the classical limit $r \rightarrow \infty$ and the integral in the
first term in Eq.\ (\ref{tempj1}) converges to $\pi/2$. This leads to
\begin{equation}
\label{intj1}
J_{1,j} = \frac{\pi}{2 \lambda} - \tau_p f(r).
\end{equation}
For typical periodic orbits $\lambda \tau_p \gg 1$ and thus we expect
the second term in (\ref{intj1}) to dominate. Since both $r$ and
$e^{\lambda \tau_p}$ are large we expect that the terms in the series
$f(r)$ are either rapidly oscillating or small. Averaging over
variations of the cut-off $r$ or a local average over $\lambda$ or
$\tau_p$ will smear out the oscillating terms. In this case the series
converges to some small value. We can conclude that $f$ is at
most of order unity, so that $J_{1,j}$ is at most of order
$\tau_p$. This suggests that the product $J_{1,1} J_{1,2}$ contributes
a term of order $\tau_p^2$ to $J$. However,
we will find that the terms proportional to $\tau_p^2$ cancel and the
integral is actually smaller.

It remains to calculate the integral of the term proportional to
$\tau_p \delta \tilde t$ in Eq.\ (\ref{factorint}),
\begin{equation}
\label{intj2}
  J_2 = r_1 r_2
  \int_0^1 du_1 du_2 \cos (r_1 u_1) \cos (r_2 u_2) \tau_p \delta \tilde{t}.
\end{equation}
Again, we separate the integration domain into regions where the
integrand is continuous. Rescaling the integration variables as
$u_j=y_je^{- m \lambda \tau_p}$, one finds
\begin{widetext}
\begin{eqnarray}
\label{scaledj2}
J_2 &=& - \frac{r_1 r_2}{\lambda}
  \sum_{l,m=0}^\infty \tau_p e^{-(m+l) \lambda \tau_p} \left\{
\int_{e^{-\lambda \tau_p}}^1 dy_1 \int_{e^{-\lambda \tau_p}}^{y_1}
dy_2 \cos \left( r_1 e^{-m\lambda \tau_p} y_1 \right) \cos \left( r_2
e^{-l\lambda \tau_p} y_2 \right)  \ln y_1 \right.
  \nonumber \\ && \left. \mbox{}
 + \int_{e^{-\lambda \tau_p}}^1 dy_1 \int_{y_1}^1 dy_2 \cos \left( r_1 e^{-m\lambda \tau_p} y_1 \right) \cos \left( r_2 e^{-l\lambda \tau_p} y_2 \right) \ln y_2 \right\}.
\end{eqnarray}
Here we used two a priori different cut-offs $r_1$ and $r_2$ for the
two integrations, instead of the single cut-off $r = c^2/\hbar$. The
use of two different cut-offs is needed to exclude products of
oscillating functions of $r_1$ and $r_2$, which would give a
non-oscillating contribution to the integral if the same cut-off would
have been used.
The integral over $y_2$ in the first term of (\ref{scaledj2}) can be
easily done. Similarly, the integral over $y_1$ can be done in the
second term, with the result
\begin{eqnarray}
  J_2 &=& - \frac{r_1}{\lambda}
  \sum_{m,l=0}^\infty \tau_p  e^{-m \lambda \tau_p} \int_{e^{-\lambda \tau_p}}^1 dy
\cos \left( r_1 e^{-m \lambda \tau_p}y \right) \left[ \sin \left( r_2 e^{-l \lambda \tau_p}y \right)
- \sin \left( r_2 e^{-(l+1) \lambda \tau_p} \right) \right] \ln y
  \nonumber \\ && \mbox{}
 - \frac{r_2}{\lambda} \sum_{m,l=0}^\infty \tau_p  e^{-l \lambda \tau_p} \int_{e^{-\lambda \tau_p}}^1 dy
\cos \left( r_2 e^{-l \lambda \tau_p}y \right)  \left[ \sin \left( r_1 e^{-m \lambda \tau_p}y \right)
- \sin \left( r_1 e^{-(m+1) \lambda \tau_p} \right) \right] \ln y.
\end{eqnarray}
Partial integration results in
\begin{eqnarray}
  J_2 &=& \sum_{l,m=0}^\infty \tau_p
  \int_{e^{- \lambda \tau_p}}^1 \frac{dy}{\lambda y}
  \sin \left(r_1 e^{-m \lambda \tau_p} y \right) \sin \left(r_2 e^{-l
  \lambda \tau_p} y \right)
  + \tau_p^2 f(r_1) f(r_2) - \frac{\pi \tau_p}{2 \lambda}
  \left[ f(r_1) + f(r_2) \right],
  \label{eq:J2final}
\end{eqnarray}
where the function $f(r)$ was defined in Eq.\ (\ref{eq:fr}) above.

In order to estimate the series of integrals in Eq. (\ref{eq:J2final}) it is
useful to divide the double sum into parts using
$$
  \sum_{l,m=0}^\infty = \sum_{l=m=0}^\infty + \sum_{m=0}^\infty
  \sum_{l=m+1}^\infty  + \sum_{l=0}^\infty \sum_{m=l+1}^\infty,
$$
and consider each part separately. The diagonal sum gives
\begin{eqnarray}
  \frac{1}{\lambda} \sum_{m=0}^\infty \int_{e^{-\lambda \tau_p}}^1
  \frac{dy}{y}
  \sin \left( r_1 e^{-m \lambda \tau_p} y \right) \left( r_2 e^{-m
  \lambda \tau_p} y \right) &=&
  \frac{1}{\lambda} \int_0^1 \frac{du}{u} \sin (r_1u) \sin (r_2 u)
  \nonumber \\ & \rightarrow &
  \frac{1}{2 \lambda} \ln \left| \frac{r_1+r_2}{r_1-r_2} \right|.
\end{eqnarray}
The fact that this integral diverges for $r_1=r_2$ is an artifact of
choosing the same cut-off for both encounters. When the cutoffs differ
the resulting expression is finite and of order
$1/\lambda$. (Certainly, there are no factors of $\tau_p$.) For the
summation with $l>m$ we find
\begin{eqnarray}
\frac{1}{\lambda} \sum_{m=0}^\infty \sum_{k=1}^\infty \int_{e^{-\lambda \tau_p}}^1 \frac{dy}{y} \sin \left( r_1 e^{-m\lambda \tau_p} y \right)
\sin \left( r_2 e^{-(m+k)\lambda \tau_p} y \right)\
=  \frac{1}{\lambda} \sum_{k=1}^\infty \int_0^1 \frac{du}{u} \sin (r_1 u)
\sin \left( r_2 e^{-k\lambda \tau_p} u \right).
\end{eqnarray}
Note that while the argument in the first sine is typically large this may not be
the case for the second sine. Let us denote by $k_0$ the point where
the exponential balances $r_2$, that is, $r_2 e^{-k_0 \lambda \tau_p}=1$.
Typically $k_0$ is not an integer and $r_2 e^{-[k_0] \lambda \tau_p} \gg 1$
while $r_2 e^{-([k_0]+1) \lambda \tau_p} \ll 1$.
For $k<k_0$ both sines are rapidly oscillating and the integral can be
approximated by integrating up to infinity.
This result in the series
\begin{eqnarray}
  \frac{1}{\lambda} \sum_{k=1}^{[k_0]} \int_0^\infty \frac{du}{u}
\sin (r_1u) \sin \left( r_2 e^{-k\lambda \tau_p} u \right)
\simeq \frac{1}{2\lambda} \sum_{k=1}^{[k_0]} \ln \left|\frac{r_1+r_2 e^{-k\lambda \tau_p}}{r_1-r_2 e^{-k\lambda \tau_p}} \right| \simeq \frac{1}{\lambda}
e^{-\lambda \tau_p}.
\end{eqnarray}
For $k>k_0$ we can expand the second sine since its argument is much
smaller than unity everywhere in the integration range. As a result
one finds
\begin{eqnarray}
\frac{1}{\lambda} \sum_{k=[k_0]+1}^\infty r_2 e^{- k \lambda \tau_p} \int_0^1
\sin (r_1u) du
&\simeq& \frac{1}{\lambda} (1-\cos r_1) \frac{e^{- ([k_0]+1))\lambda \tau_p}}{1-e^{-\lambda \tau_p}}
\end{eqnarray}
\end{widetext}
This part of the sum is much smaller than the other contributions. The case
where $m>l$ is treated in the same way. Thus, we can conclude that the
series of integrals in Eq.\ (\ref{eq:J2final}) is of typical size
$\tau_p/\lambda$.

Combining the integrals $J_1$ and $J_2$ we find that $J_2 - J_{1,1}
J_{1,2}$ is of order $\tau_p/\lambda$. Since it is in this combination
that these integrals enter into the Eq.\ (\ref{factorint}) for $J$,
we conclude that $J$ is of order $\tau_{\rm D}/\lambda$, as
advertised.

\section{Time reversal symmetry}
\label{TRS}

\begin{figure}[t]
\epsfxsize=0.9\hsize
\hspace{0.01\hsize}
\epsffile{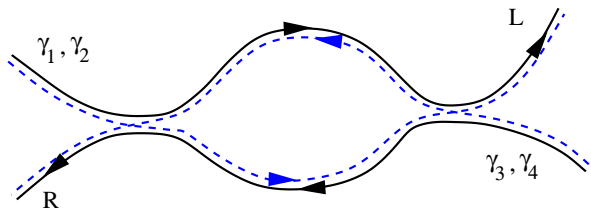}
\caption{(Color online)
  \label{trsfig} In the presence of time-reversal symmetry one
  also has to consider interference of time-reversed trajectories. The
  figure shows the configuration of type A after time-reversal of the
  trajectories $\gamma_3$ and $\gamma_4$.}
\end{figure}

The calculation of the pumping current variance performed in Sec.\
\ref{semic} did not include contributions from trajectory
configurations which depend on the presence or absence of time
reversal symmetry. Indeed, all stretches of interfering trajectories
in Fig.\ \ref{typea} are traversed in the same direction, so that any
accumulated phase difference does not depend on an applied magnetic
field. For conductance fluctuations, there is an equal contribution to
the variance of the conductance in which
one pair of trajectories is replaced by the time-reversed
trajectories, so that
the interfering stretches are traversed in opposite directions, as,
{\em e.g.,} in Fig.\ \ref{trsfig}.  (In the language of diagrammatic
perturbation theory, such contributions are referred to as
`Cooperons'.) Such contributions depend on the presence or absence of
time-reversal symmetry and are responsible for the factor-two
suppression of $\mbox{var}\, G$ upon application of an external
magnetic field.\cite{kn:altshuler1995,kn:stone1995}
For a quantum pump, however, there is no contribution
from trajectories that traverse the interference region in opposite
directions, so that $\langle I^2 \rangle$ is independent of the
presence or absence of time-reversal symmetry. This is well known in
terms of random matrix theory.\cite{kn:brouwer1998,kn:shutenko2000} A
semiclassical derivation of this fact is given below.

The contribution of type A to the correlator $\left< \Pi^L \Pi^R
 \right>$ is shown in Fig. \ref{trsfig}, but with the trajectories
 $\gamma_3$ and $\gamma_4$ replaced by their time-reversed.
The main difference between this configuration of trajectories and
configurations which do not depend on time reversal symmetry is
that the trajectories in the central loop propagate in
opposite directions. We'll now show that the contribution of such a
configuration of classical trajectories to $\left< \Pi^L \Pi^R
\right>$ is zero, even in the absence of an applied magnetic field.

We first consider the contribution of the type ``A1'', in which both
encounter regions fully reside inside the quantum dot. For this
contribution, the reversal of the direction of two of the trajectories
has no effect, and one finds that this contribution is given by Eq.\
(\ref{lasta1}) of Sec.\ \ref{semic}. For the configuration
depicted in Fig.\ \ref{trsfig} both encounters can touch the leads,
resulting in three additional contributions to $\langle \Pi^L \Pi^R
\rangle$. Each of the configurations in which only one
encounter touches the lead openings has a contribution given by Eq.\
(\ref{eq:IRLA2}) of Sec.\ \ref{semic} (up to the interchange $t_{{\rm
    enc},1} \leftrightarrow t_{{\rm enc},2}$ in one of these).
The last contribution, from
trajectory configurations in which both encounters touch the leads,
was not considered in Sec.\ \ref{semic}. In such a configuration,
$\gamma_{1}$ and $\gamma_{2}$ enter through the right contact and
$\gamma_{3}$ and $\gamma_{4}$ to enter through the left contact. The
calculation of this contribution is similar to that of the
contribution ``A2'' in Sec.\ \ref{semic}, but with integrations over
two encounter times instead of only one. Labeling this configuration
``A4'', we find
\begin{widetext}
\begin{eqnarray}
\label{trsa4}
  \left< \Pi^L \Pi^R \right>_{A4} &=& \frac{32 N_1 N_2 \tau_{\rm
    D}^3}{N^2} C_1 C_2 r^2 \lambda^2
  \int_{0}^1 dx_1 dx_2 \cos(r x_1) \cos(r x_2)
  \\ && \mbox{} \times
   \left\{ 5 \tau_{\rm D} \left( 1- x_1^{1/\lambda\tau_{\rm D}}
    \right) \left( 1- x_1^{1/\lambda\tau_{\rm D}} \right)
    -2 t_{{\rm enc},2} x_2^{1/\lambda\tau_{\rm D}} \left( 1-
    x_1^{1/\lambda/\tau_{\rm D}} \right)
    -2 t_{{\rm enc},1} x_1^{1/\lambda\tau_{\rm D}} \left( 1-
    x_2^{1/\lambda \tau_{\rm D}} \right) \right\},
  \nonumber
\end{eqnarray}
where $t_{{\rm enc},j} = (1/\lambda) \ln (1/x_j)$, $j=1,2$.
Combining all contributions we then find that the additional
contribution to the correlator $\langle \Pi^L \Pi^R \rangle$ in the
presence of time-reversal symmetry is
\begin{eqnarray}
  \label{trsa}
  \left< \Pi^L \Pi^R \right>^{{\rm TRS}}_{A} &=&
  \frac{32 N_1 N_2 \tau_{\rm D}^4}{N^2} C_1 C_2 r^2 \lambda^2 \int_0^1 dx_1 dx_2
  \cos(r x_1) \cos(r x_2)
  \left( 2  x_1^{1/\lambda\tau_{\rm D}} +
  2 x_2^{1/\lambda\tau_{\rm D}} -5 \right).
\end{eqnarray}
\end{widetext}
Upon integration over $x_1$ and $x_2$, all three terms between
brackets yield fast oscillating functions of $r = c^2/\hbar$, which
are neglected in the classical limit.
The calculation of contributions of type B proceeds along in exactly the
same way. We have verified that the resulting contribution can also be
neglected.

\end{appendix}


\end{document}